\documentclass[twocolumn,showpacs]{revtex4}
\usepackage{latexsym,amssymb,amsmath,graphicx}

\newcommand{\zerodel}{.\kern-\nulldelimiterspace}

\begin{document}

\def\pr{\prime}
\def\be{\begin{equation}}
\def\en#1{\label{#1}\end{equation}}
\def\d{\dagger}
\def\bar#1{\overline #1}
\newcommand{\per}{\mathrm{per}}

\newcommand{\rd}{\mathrm{d}}
\newcommand{\vare}{\varepsilon }

\title{Diffusive lossless energy and coherence transfer by noisy coupling}
\author{D. Mogilevtsev$^{1}$, G. Ya. Slepyan$^{2}$}

\affiliation{$^{1}$Institute of Physics, Belarus National Academy
of Sciences, F.Skarina Ave. 68, Minsk 220072 Belarus;\\
$^2$Department of Physical Electronics, School of Electrical
Engineering, Faculty of Engineering, Tel Aviv University, Tel Aviv
69978, Israel}

\begin{abstract}
Here we show that noisy coupling can lead to diffusive
lossless energy transfer between individual quantum systems
retaining a quantum character leading to entangled stationary states.
Coherence might flow diffusively while being summarily preserved even when energy exchange is absent.
Diffusive dynamics persists even in the case when additional
noise suppresses all the unitary excitation exchange:
arbitrarily strong local dephasing, while destroying quantum correlations,
is not affecting energy transfer.
\end{abstract}

 \maketitle

\textit{Introduction} $\quad$ Diffusive transfer of energy (and,
ultimately, derivation of Fourier heat-transfer law from
microscopic dynamics) up to day remain subjects of theoretical
interest and even controversy
\cite{lepri,mahler,kosloff,bonetto,dhar}. For microscopic
dynamics dominated by quantum effects establishing of diffusive
energy transfer is far from being obvious. Commonly
considered microscopic models, such as chains of unitarily
connected networks of bosonic and/or fermionic systems with
attached thermal reservoirs and noise sources can demonstrate both
ballistic and diffusive behavior in dependence on interaction
strengths and other parameters of the whole system, generally
requiring approximations (such as long time and large size limits)
for emergence of classical-like heat dynamics. Here we suggest a
noise-mediated microscopic mechanism for diffusive transfer. Energy can
propagate without loss, but through losses.
Recently it has become quite usual to see noises not
only as something destroying quantum coherence and reducing
quantum states to classicality, but also as a tool to create and
enhance quantumness. Non-local loss can preserve entanglement and
even generate entangled stated from initially uncorrelated ones
\cite{ekert,zanardi,braun,benatti,lidar}. Engineered loss can lead to
dissipative protection and coherence preservation \cite{dav2001,man1,man2} and deterministic creation
of non-classical states
\cite{zoll96,TFAb,TFComm,TFRepl,parkins2003} and serve as a tool
for quantum computing \cite{cirac2009,cirac2011}. Networks of
dissipatively coupled systems can support topologically protected
states \cite{zollertop}. Even a pure local dephasing is no longer
considered completely harmful: it can enhance quantum state transfer
and suppress localizing effects of static disorder \cite{plenio2010,silbey,moix}.
However, too strong local dephasing generally suppresses unitary excitation exchange and
energy transfer stemming from it.

Here we show a microscopic mechanism of  diffusive
lossless energy transfer, which is impervious to local dephasing.
It arises when coupling constants describing common single-excitation hopping are
fluctuating randomly, like it is, for example, with spin-spin dipolar coupling in random environments.
Dynamics produced by fluctuating coupling  might preserve certain quantum
correlations, and even entanglement during evolution
toward the stationary state. Populations are not coupled by the
dynamics with the off-diagonal terms. So, for example, a diagonal
initial state evolves to a symmetrical mixture of diagonal states,
whereas coherences (i.e., off-diagonal elements or superpositions of them) can also flow diffusively, and the sum of
certain coherences can be preserved by dynamics.
Diffusive coherence flow can occur without energy exchange.
Such a dynamics can occur for different quantum systems, for example, two-level systems and bosonic modes.  The latter case is remarkable. It is known that the light in a structured surrounding (such as photonic crystals) can produce coherent dynamics typical rather for charged or neutral particles, but not for photons, such as  Blokh oscillations with single photons in the waveguide lattices \cite{rai,moran,pert}, Rabi-oscillations of photons \cite{shandar} and effective magnetic field for photons by controlling the phase of dynamic modulation\cite{fang}. Here we have one more effect: photon diffusion without energy losses due to noisy coupling.

\begin{figure}[htb]

\includegraphics[width=0.75\linewidth]{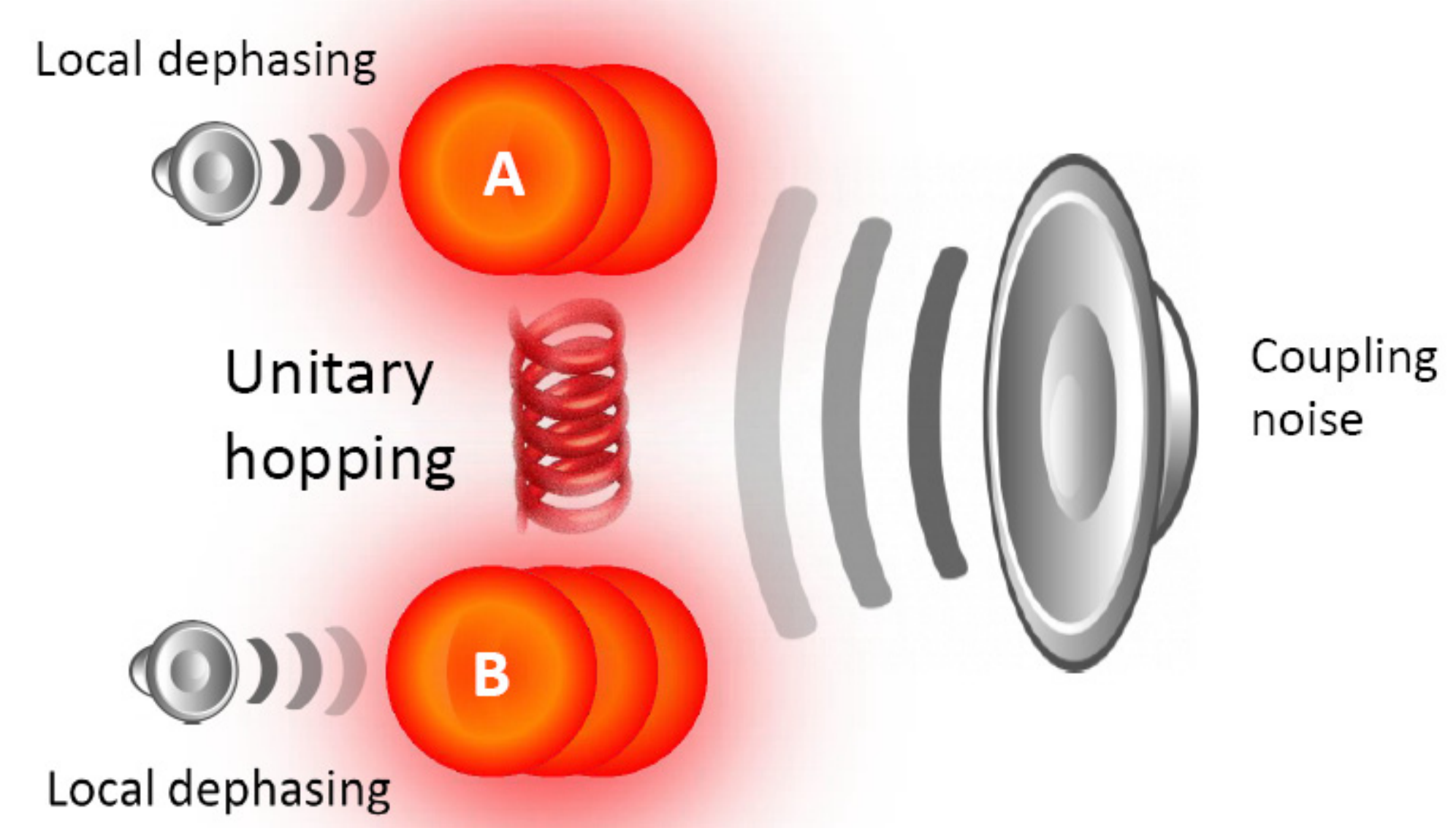}

\caption{(color online) A schematic depiction of the considered scheme with possible local dephasing noises
affecting transition frequencies, possible non-fluctuating unitary hopping and non-local coupling noise (described by variables
$\eta_j(t)$ in Eq.(\ref{basic h1})) affecting both unitary hopping rate
and transition frequencies.} \label{fig1}
\end{figure}

\textit{Simple model} $\quad$ To show the essence of our
diffusive transfer mechanism, let us start with the simple
illustrative model. We consider a tight-binding chain of identical
two-level systems (TLS) with the following interaction Hamiltonian
\begin{equation}
H=\hbar
\sum_{j=1}^N\eta_j(t)(\sigma^+_j+\sigma_{j+1}^+)(\sigma^-_j+\sigma_{j+1}^-),
\label{basic h1}
\end{equation}
where the operators $\sigma^{\pm}_j=|\pm_j\rangle\langle\mp_j|$;
the vector $|\pm_j\rangle$ describes the upper (+) or lower
(-) levels of $j$-th TLS (see Fig.\ref{fig1}). Quantities $\eta_j(t)$ describe
classical real zero-mean independent white noises, $\langle
\eta_j(t)\rangle=0$, $\langle
\eta_j(t)\eta_k(\tau)\rangle=\gamma_j\delta_{jk}\delta(t-\tau)$;
quantities $\gamma_j\geq 0$, $\forall j$ we term "hopping diffusion rates". Physically,
the model (\ref{basic h1}) corresponds to the chain with the same
noise affecting only two neighbors with the energy levels and
interaction strengths perturbed in the same way\cite{notice1}.
Deriving the
master equation in the standard way \cite{BP}, one gets from
Eq.(\ref{basic h1}) the following master equation
\begin{equation}
\frac{d}{dt}\rho=2\sum\limits_j\gamma_j \left(L_j\rho
L_j-L_j\rho-\rho L_j\right), \label{master1}
\end{equation}
with $L_j=(\sigma^+_j+\sigma_{j+1}^+)(\sigma^-_j+\sigma_{j+1}^-)$.
It is obvious from Eq.(\ref{master1}) that the sum of TLS upper-level populations (which we further address as energy),
$E_0=\sum\limits_{j=1}^{N+1}n_j$, where
$n_j=\langle\sigma^+_j\sigma^-_j\rangle$, is preserved.
Eq.(\ref{master1}) leads directly to the diffusive transfer
equation for individual populations
\begin{eqnarray}
\frac{d}{dt}n_j=-2(\gamma_j+\gamma_{j-1})n_j+
2\gamma_{j}n_{j+1}+2\gamma_{j-1}n_{j-1}.
 \label{dif1}
\end{eqnarray}
Further, assuming the chain homogeneous, we take $\gamma_{j}\equiv\gamma$ for $1\leq j\leq N$ and $\gamma_{j}=0$ for $j\leq 0,j\geq N+1$.
it follows from Eq.(\ref{dif1}) that in the long time limit the equilibrium is reached, $n_j\rightarrow n_{st}=E_0/(N+1)$,
$\forall j$. Introducing a sum of local energies from $k$th to $l$th TLS,
$E_{k,l}(t)=\sum\limits_{j=k}^ln_j(t)$, $k,l\neq 1,N+1$,
from Eq.(\ref{dif1}) it follows that
$\frac{d}{dt}E_{k,l}=S_{l+1}-S_{k-1}$, with the local flux defined as
$S_k=2(n_{k+1}-n_{k})$. The energy balance in a region of
the chain is naturally defined by the energy flowing through the
borders.

\begin{figure}[htb]

\includegraphics[width=\linewidth]{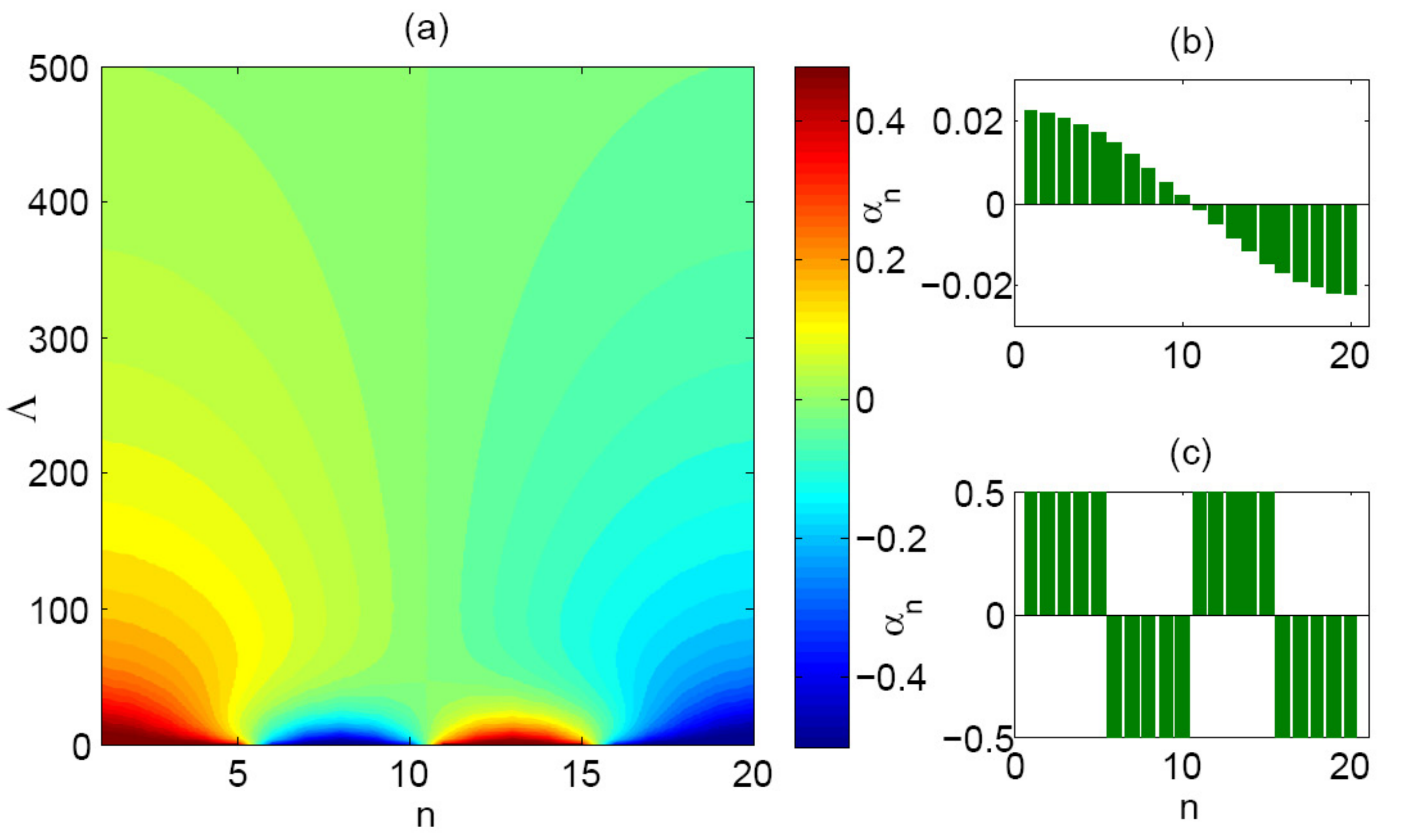}

\caption{(color online)(a)Dynamics of coherences, $\alpha_n$, as given by Eq.(\ref{master1}) for the energy-equilibrated initial state (\ref{noheat}) with initial distribution of $\alpha_n$ shown in the panel (c). The panel (b) shows distribution of  $\alpha_n$ for $\Lambda=500$, $\Lambda=\gamma t$.} \label{fig2}
\end{figure}

\textit{Non-classicality} $\quad$ Diffusive energy transfer does
not imply loss of quantum correlations of the chain state.
Let us demonstrate it with an example of single-excitation
dynamics. Firstly, the system described by Eq.(\ref{master1})
has an entangled stationary state
$|\psi_{st}\rangle=\sum\limits_{j=1}^{N+1}(-1)^j|1_j\rangle/\sqrt{N+1}$,
satisfying $L_j|\psi_{st}\rangle=0$, $\forall j$, where the vector
$|1_j\rangle=|+_j\rangle\prod\limits_{k\neq j}|-_k>$. Similar
entangled stationary states were found recently in dissipatively
coupled TLS chains \cite{our2015}. Secondly, coherences in the
chain can  also flow diffusively having the sum of them preserved. For example, assuming the
single-excitation initial state, for coherences defined as,
$\alpha_k=(-1)^k\langle 1_k|\rho|0\rangle$, one has equation formally coinciding with the equation for a classical random walk
\[\frac{d}{dt}\alpha_k=2\gamma(-2\alpha_k+
\alpha_{k+1}+\alpha_{k-1})\] for $1<k<N+1$, where $|0\rangle$ is
the vector describing all the TLS of the chain being in the lower
level. Curiously, the coherence might flow even if the energy
gradient is absent. Indeed, this will take place, for example, for
the initial state being a mixture of phase states of each TLS:
\begin{equation}
\rho(0)=\sum\limits_{j=1}^{N+1}|\theta_j\rangle\langle
\theta_j|/(N+1),
\label{noheat}
\end{equation}
where
$|\theta_j\rangle=(|+_j\rangle+\exp{i\theta_j}|-_j\rangle)\prod\limits_{k\neq
j}|-_k>/\sqrt{2}$, and the angles $\theta_j$ are arbitrary. In the
long-time limit the state will be the following mixture
$\rho(\infty)=\sum\limits_{j=1}^{N+1}\rho_j$, where
$\rho_j=({\mathbf{1}}_j+(-1)^j(\Theta^{\ast}\sigma_j^++\Theta\sigma_j^-))\otimes\prod\limits_{k\neq
j}|-_k\rangle\langle-_k|/2(N+1)$, where
$\Theta=\sum\limits_{j=1}^{N+1}(-1)^j\exp\{i\theta_j\}$.
This situation is illustrated in Fig.\ref{fig2}, where it is shown how evolves the state (\ref{noheat})
with $\theta_j$ equal to $0$ or $\pi$. Coherences, $\alpha_n$, oscillate between positive and negative values. However, energy exchange is not taking place at all times. For this example $\Theta=0$, so, asymptotically the state of the chain is the diagonal mixture of the vacuum and single-excitation states. For long times, $4\gamma t\sin^2\{\pi/(N+1)\}\gg 1$, phase of chain halves are opposite, since the initial state depicted in Fig.\ref{fig2}(c) has non-zero overlap with the simplest antisymmetric eigenmode of the chain \cite{our2015}.
For
coherences $\alpha_{kl}=(-1)^{k+l}\langle 1_k|\rho|1_l\rangle$ and
$k\neq l\pm 1$ Eq.(\ref{master1}) for $1<k,l<N+1$ one obtains equations for 2D classical random walk:
\[\frac{d}{2\gamma dt}\alpha_{kl}=-4\alpha_{kl}+
\alpha_{k+1,l}+\alpha_{k-1,l}+\alpha_{k,l+1}+\alpha_{k,l-1}.\] For
$k=l\pm 1$ one has equations not coinciding with ones for 2D classical random walk
\[\frac{d}{2\gamma
dt}\alpha_{k,k+1}=-2\alpha_{k,k+1}+\alpha_{k-1,k+1}+\alpha_{k,k+2}.\]
It is seen from these equations that quantities $\alpha_{kl}$ for
$k\neq l$ are not coupled with populations, $\rho_{kk}$. However,
they are not arbitrary. The matrix $\rho(t)$ should always be semi-positive, so, for example,
one has $|\alpha_{kl}(t)|\leq\sqrt{\rho_{kk}(t)\rho_{ll}(t)}$ for any $t$. Absence of
coupling between diagonal and off-diagonal elements of the density
matrix leads to preservation of the diagonality of the state.

\textit{Symmetrization} $\quad$
As follows from Eq.(\ref{master1}), a single initial excitation leads asymptotically to the
symmetrical mixture
$\rho(\infty)=\sum\limits_{j=1}^{N+1}|1_j\rangle\langle
1_j|/(N+1)$.
Now let us demonstrate that the
symmetrization takes place for an arbitrary initial
diagonal state. Indeed, for the variable $n(\mathcal{K})=\langle
\sigma^+_{j_1}(t)\sigma^-_{j_1}(t)\ldots\sigma^+_{j_K}(t)\sigma^-_{j_K}(t)\rangle$,
where $\forall j_k\in \mathcal K$ one has from Eq.(\ref{master1})
that
\begin{eqnarray}
\nonumber
\frac{d}{dt}n(\mathcal{K})=-2\sum\limits_{j\in\mathcal{K}}
\gamma_j\varsigma_{j+1}(n(\mathcal{K})-n(\mathcal{K}:j\rightarrow
j+1))-\\
-2\sum\limits_{j\in\mathcal{K}}
\gamma_{j-1}\varsigma_{j-1}(n(\mathcal{K})-n(\mathcal{K}:j\rightarrow
j-1)),
 \label{dif2}
\end{eqnarray}
where the set $(\mathcal{K}:j\rightarrow k)$ is the set
$\mathcal{K}$ with $j$-th TLS replaced with $k$-th; the
coefficients $\varsigma_{k}=0$  for $k\in\mathcal{K}$ and are
unity otherwise. Eq. (\ref{dif2}) leads to a number
of interesting consequences. For examples, not only excitations,
but also the absence of excitations can propagate diffusively. Let
us consider vectors $|0_j\rangle=|-_j\rangle\prod\limits_{k\neq
j}|+_k\rangle$. It follows from Eq.(\ref{dif2}) that the
probability, $n_{j}=\langle0_j|\rho|0_j\rangle$ of non-excitation
of the chain TLS satisfies the 1D diffusion equation (\ref{dif1}).
Just like it was for the single-excitation case, off-diagonal
elements do not couple to the diagonal ones.  But the most
importantly, the diffusion equation (\ref{dif2}) shows that any
initial diagonal state $\rho(0)=\sum_{n=0}^{N+1}\rho_n$,
with energy $E=\sum_{n=0}^{N+1}\mathrm{Tr}\{\rho_n\}n$,
where $\rho_n$ are states with exactly $n$ excited TLS, is transformed to the symmetrical state
\begin{eqnarray}
\rho(0)\rightarrow\rho(\infty)=\sum_{n=0}^{N+1}\mathrm{Tr}\{\rho_n\}S_{n,N+1},
 \label{sym1}
\end{eqnarray}
which the same energy, where $S_{n,N+1}$ is the phase-averaged Dicke state being an equal
mixture of projectors of all permutations of $n$ excited TLS from
the total number, $N+1$, of them \cite{dicke}.

\textit{Thermalization} $\quad$ Dynamics described by  Eqs.
(\ref{dif1},\ref{dif2}) preserves the number of excitation and does not lead to the global Gibbs state of
the chain. However, local populations, $n_j$, still can be given a
meaning of the local temperature at least for states close to the
stationary one (\ref{sym1}). Firstly, any single-particle state
obtained by averaging over $N$ TLS of the state (\ref{sym1}), is
the Gibbs state,
$\rho_j(\infty)=\exp\{-\beta\sigma^+_{j}\sigma^-_{j}\}/\mathrm{Tr}(\exp\{-\beta\sigma^+_{j}\sigma^-_{j}\})$:
where the parameter $\beta=-ln\{(1-{\bar n })/{\bar n }\}$, and
${\bar n }=\sum_{n=0}^{N+1}\mathrm{Tr}\{\rho_n\}n/(N+1)$ being the
population of each TLS. Then, the state (\ref{sym1}) is the
typical one for which the canonical typicality takes place
\cite{zanghi,popesku}. Namely, for small subset of $m\ll N+1$ TLS
the state of these $m$ TLS averaged over the rest of the chain, is
very close to the to the Gibbs state $\rho_m^{Gibbs}=\exp\{-\beta
H_m\}/\mathrm{Tr}(\exp\{-\beta H_m\})$ with $H_m$ is the sum of
$\sigma^+_{j}\sigma^-_{j}$ of the subset. For TLS and mixed state
(\ref{sym1}) there is the exact bound,
$||\rho_m(\infty)-\rho_m^{Gibbs}||_1\leq 4m/(N+1)$, where
$\rho_m(\infty)$ is the state of the subset with $m$ TLS averaged
over the rest of the chain \cite{muller}. Thus, for the
state of the chain weakly deviating from the symmetric one
(\ref{sym1}), one can meaningfully introduce the temperature
as $T=\hbar\omega/{\mathrm k}\beta$, and derive the continuous heat-transfer
equation.  Let us do it as outlined in Ref.\cite{segal}.  For the heat flux between the
neighbor TLS in the continuous limit one has
$J=2\gamma_j\hbar\omega(n_{j+1}-n_j)\rightarrow -2C_V(T)a\gamma(x)
\frac{\partial}{\partial x}T$, where $a$ is the distance between
the neighbor TLS, and the specific heat $C_V={\partial u
}/{\partial T}$, and the local internal energy is
$u(x)=\hbar\omega n(T(x))$. Thus, one gets the following Fourier
equation
\begin{eqnarray}
\frac{\partial}{\partial t}u\approx- \frac{\partial}{\partial x}
\left( \kappa(x,T)\frac{\partial}{\partial
x}T\right),\label{heat2}
\end{eqnarray}
where the thermal conductivity $\kappa(x,T)=2a^2\gamma(x)C_V(T)$.
The specific heat for the chain is $C_V={\mathrm
k}\beta^2e^{\beta}(e^{\beta}+1)^{-2}$. The dependence of the heat
conductivity of the temperature is defined by the hopping diffusion rate,
$\gamma$. For the rate independent of the
chain temperature,
one has a common $T^{-2}$ dependence \cite{peierls}.

\textit{Generalizations and realizations} $\quad$ Emergence of the diffusive lossless
energy transfer through coupling noise is quite general phenomenon not restricted to TLS systems.
For a single-excitation case the dynamics is completely similar for TLS chain or bosonic one, or even the system of coupled TLS and modes.
Obviously, any unitary hopping term preserving the number of excitations and with zero-mean random interaction constants under condition of the Markovian approximation applicability would lead to diffusive energy transfer.
For the chain of bosonic modes the hopping term is
\begin{equation}
H=\hbar
\sum_{j=1}^N\eta_j(t)(a^+_j+a_{j+1}^+)(a_j+a_{j+1}),
\label{basic hb1}
\end{equation}
with bosonic creation, $a^+_j$, and annihilation, $a_j$ operators satisfying $[a_j,a^+_k]=\delta_{jk}$. It is easy to see that for
the independent white noises, $\eta_j(t)$, one gets the standard Lindblad master equation with $L_j=(a^+_j+a_{j+1}^+)(a_j+a_{j+1})$,
which leads directly to the diffusive transfer
equation (\ref{dif1}) for the modal average number of photons, $n_j=\langle
a^+_ja_j\rangle$. Just like the TLS chain, diagonal and off-diagonal elements of the total chain density matrix in the energy basis are not coupled by dynamics. Also, the initial diagonal states remain diagonal and are asymptotically symmetrized.

Bosonic scheme offers the simplest way to realize, test and use the discussed diffusive transfer scheme. For example, the system of coupled waveguides similar to ones recently used for  demonstration of localized states in ideal defectless Lieb lattices can be used for a purpose \cite{erica,vic}. Modulation of waveguide coupling constants achieved by random variation of distance, waveguide dimension and/or dielectric constant of the bulk can be implemented. In this way, it is possible realizing a lossless optical equalizer for suppressing both intensity and phase
fluctuations of multi-mode fields, which is currently a
topical problem \cite{pashotta} Another possible realizations one can find in schemes for dynamical suppression of decoherence \cite{cur}, in arrays of  persistent-current Josephson qubits \cite{mooij} with dynamical coupling \cite{lyahov}, or the chain of nitrogen-vacancy centers in diamond in fluctuating magnetic field \cite{zelezko}.

\textit{Local dephasing and unitary hopping} $\quad$ In any realistic schemes involving random uncontrolled variation of coupling constants, one naturally expects having local dephasing. Also, when fluctuating coupling constants are not of zero mean, one should expect unitary hopping.
To consider an influence of these phenomena on the energy transfer, let us again consider the chain of TLS.  We take the interaction
Hamiltonian as
\begin{eqnarray}
\nonumber
\frac{H}{\hbar}=\sum_{j=1}^{N+1}R_j(t)\sigma^+_j\sigma_{j}^- +
\sum_{j=1}^{N}G_j(t)(\sigma^+_j\sigma_{j}^-+\sigma^+_{j+1}\sigma_{j+1}^-)+\\
\sum_{j=1}^N ((g_j+v_jG_j(t))\sigma^+_j\sigma_{j+1}^-
+(g_j^{\ast}+v_j^{\ast}G_j(t))\sigma_{j+1}^+\sigma^-_j),
 \label{h2}
\end{eqnarray}
where constants $g_j$ describe the excitation exchange strengths
between the neighbor TLS; the constants $v_j$ describe strengths
of TLS interaction with the corresponding reservoir.  Operators $R_j(t)$ are describing local dephasing;
operators $G_j(t)$ describe common reservoirs for neighbor TLS. We
assume the Markovian limit taking $\langle
X_j(t)Y_k(\tau)\rangle=\gamma^{X,Y}_j\delta_{XY}\delta_{jk}\delta(t-\tau)$
with $X,Y=R,G$. Again, the standard derivation procedure leads
from the Hamiltonian (\ref{h2}) to the following master equation
\begin{eqnarray}
\nonumber
\frac{d}{dt}\rho=i[V,\rho]+\sum\limits_{j=1}^{N+1}\gamma_j^{R}
\left(2L_j^R\rho L_j^R-L_j^R\rho-\rho L_j^R\right)+ \\
\sum\limits_{j=1}^{N}\gamma_j^{G} \left(2L_j^G\rho
L_j^G-(L_j^G)^2\rho-\rho (L_j^G)^2\right), \label{master2}
\end{eqnarray}
where $L_j^R=\sigma_{j}^+\sigma_{j}^-$,
$L_j^G=\sigma^+_j\sigma_{j}^-+\sigma^+_{j+1}\sigma_{j+1}^-+v_j\sigma^+_j\sigma_{j+1}^-
+v_j^{\ast}\sigma_{j+1}^+\sigma^-_j$, and the unitary part is
$V=\sum\limits_{j=1}^N(g_j\sigma^+_j\sigma_{j+1}^-+
g_j^{\ast}\sigma_{j+1}^+\sigma^-_j)$. Deriving equations for
average populations, from Eq.(\ref{master2}) one gets
\begin{eqnarray}
\nonumber \frac{d}{dt}n_j=i\sum\limits_{k=j-1,j}\langle
g_{k}\sigma^+_k\sigma_{k+1}^--g_k^{\ast}\sigma_{k+1}^+\sigma^-_k\rangle + \\
\nonumber
2(\gamma_j|v_j|^2+\gamma_{j-1}|v_{j-1}|^2)n_j- \\
2\gamma_{j}|v_{j}|^2n_{j+1}-2\gamma_{j-1}|v_{j-1}|^2n_{j-1}.
 \label{dif3}
\end{eqnarray}
It is immediately seen from Eq.(\ref{dif3}) that in absence of unitary exchange, $g_k=0$, $\forall j$, localized dephasing does not influence at all the diffusive transfer caused by fluctuating coupling constants. The unitary hopping is more harmful,
since it couples diagonal and off-diagonal matrix elements. However, in this case local dephasing can actually save the day.
Let us assume that localized dephasing is by far the strongest
factor influencing dynamics of TLS (as it natural for realistic
noisy structures and larger temperatures, when the Markovian limit
holds for dephasing \cite{BP}), $\gamma_j^{R}\gg
\gamma_k^{R},|v_k|^2\gamma_k^{R}, |g_k|$, $\forall j,k$. Then, it
is easy to get from Eq.(\ref{dif3}) that
$\langle\sigma^+_k(t)\sigma_{k+1}^-(t)\rangle\approx
\langle\sigma^+_k(0)\sigma_{k+1}^-(0)\rangle\exp\{-(\gamma^R_k+\gamma^R_{k+1})t\}+O([\min\limits_j\gamma_j^R]^{-1})$.
Strong local dephasing suppresses unitary excitation exchange.
However, this dephasing does not affect energy transfer produced
by the correlated dephasing. Thus, for times much exceeding $\max\{1/\gamma_j^{R}\}$ and hopping diffusion rates much exceeding $\max\{|g_j|^2/\gamma_j^{R}\}$ the first term on the
right-hand part of Eq.(\ref{dif3}) can be neglected, and populations, $n_j(t)$
change diffusively as described by Eq.(\ref{dif1}). Due to possibility of neglecting the unitary hopping, also correlation functions $n(\mathcal{K})$ would evolve according to Eq.(\ref{dif2}) with the final state being symmetrized and thermalized, as
it was described by Eq.(\ref{sym1}).

\textit{Conclusions} $\quad$ We have discussed microscopic mechanism that leads to diffusive
lossless energy transfer on the level of few quanta. We have shown that noise of unitary hopping constants under the conditions of the Markovian approximation applicability leads to the energy diffusion in a tight-binding systems of quantum systems of different nature, be it, for example, spins or field modes. In absence of additional local dephasing, the stationary state can be entangled. Local dephasing, even strong to such extent that unitary hopping is suppressed, does not break the dynamics leading to complete symmetrization of the stationary state. For sufficiently large systems, locally such a stationary state is very close to the Gibbs state.  Thus, one can introduce a temperature in a standard way and derive the heat-transfer equation. The dynamics was considered for 1D chain, but there are obvious generalizations for 2D and 3D ones. We have suggested practical systems were the scheme can be verified experimentally: the sets of coupled waveguides with fluctuating coupling, dynamically controlled superconducting qubits, color centers in diamonds.The suggested mechanism of diffusive lossless energy transfer can be responsible for energy transfer in strongly noised coupled quantum systems (such as biomolecules) at high temperature.

D.M. thankfully acknowledges the support of the European Commission through the SUPERTWIN project, id.686731,  G. Ya. acknoledges the supports by the EU Horizon 2020 project H2020-MSCA-RISE-2014-644076 CoExAN and EU FP7 projects FP7-PEOPLE-2009-IRSES-247007 CACOMEL, FP7-PEOPLE-2009-IRSES-246784 SPINMET, FP7-PEOPLE-2012-IRSES-316432 QOCaN and FP7-PEOPLE-2013-IRSES-612285 CANTOR. D.M. and G. Ya. thanks A.P. Nizovtsev for fruitful discussions and pointing to relevant references.


\begin{thebibliography}{99}
\bibitem{notice1} Notice that
diagonal terms in the Hamiltonian (\ref{basic h1}) do not change equations for diagonal elements of the density matrix, and do not lead to coupling between diagonal and off-diagonal elements. However, we retain this terms, firstly, for making obvious an existence of entangled stationary state, and, secondly, to make equation for coherences between zero and single excitation subspace look like a well-known 1D equation for a classical random walk (discretized heat-transfer equation), which makes obvious preservation of the sum of coherences.
\end{thebibliography}
\end{document}